\documentclass[seceq]{ptptex}

\usepackage{wrapft}
\usepackage{amssymb}
\usepackage{here}
\usepackage[dvips]{graphicx}

\title{
 Reconsideration of the 2-flavor NJL model with dimensional 
 regularization at finite temperature and density
}

\author{
 Takahiro \textsc{Fujihara},$^1$ Tomohiro \textsc{Inagaki},$^2$
 Daiji \textsc{Kimura},$^1$ and 
 Alexander \textsc{Kvinikhidze}$^3$
}

\inst{
    $^{1}$Department of Physics, Hiroshima University,
 Higashi-Hiroshima, \\
 739-8526, Japan \\
 $^2$Information Media Center, Hiroshima University,
 Higashi-Hiroshima, \\
 739-8521, Japan, \\ 
 $^3$A. Razmadze Mathematical Institute of Georgian Academy of Sciences,\\
 M. Alexidze Str. 1, 380093 Tbilisi, Georgia
}

\abst{
We study the 2-flavor Nambu--Jona-Lasinio (NJL) model at finite
temperature, $T$, and chemical potential, $\mu$,  in the dimensional 
regularization. The pion and sigma meson masses are calculated at 
finite $T$ and $\mu$. The color superconductivity is also discussed 
in the extended NJL model. We obtain the phase structure of the chiral 
and color symmetry in the dimensional regularization and
compare the results with ones obtained  in the cut-off regularization.
}

\begin{document}

\maketitle

\section{Introduction}
The asymptotic freedom of QCD and consequently perturbation theory 
unfortunately is of no help in the study of the low energy phenomena 
of QCD, therefore effective models of QCD are widely used. Here we 
regard the NJL model as a low energy effective model of QCD and evaluate 
the phase diagram at finite temperature and chemical potential. NJL model 
contains a four-fermion interaction corresponding to a dimension 6
operator in the Lagrangian. Thus the model is non-normalizable in four 
space-time dimensions and some regularization should be used. It is not 
surprising that the results may depend on the regularization scheme.
One usually introduces the cut-off scale to regularize the model. The 
scale is fixed to reproduce the pion physics. Unfortunately the Fermi 
momentum exceeds the cut-off parameter in the color superconducting phase.

Schwinger-Dyson (SD) equation is a commonly used means to consider 
non-perturbative effects in QCD. One can evaluate the dynamically 
generated fermion (quark) mass by using SD equation for fermion
propagator. It is known that SD equation gives 
similar results to NJL model in two dimensions. If we apply 
the ladder approximation, the instantaneous exchange 
approximation and neglect momentum dependent parts of the 
fermion self-energy in SD equation, the result coincides 
with the gap equation in two dimensional NJL model at the 
leading order of $1/N_c$ expansion.
Two dimensional NJL model is renormalizable, but too simple
for our proposes, therefore here we consider the model in 
$D$ ($2 < D < 4$) dimensions.

\section{Meson masses}

The two-flavor NJL model is defined by \cite{njl,njl2} 
\begin{eqnarray}
 \mathcal{L}_{\rm NJL} = \bar{\psi}(i \partial\!\!\!/ -m) \psi
    + g\{(\bar{\psi}\psi)^2
    + (\bar{\psi}i\gamma_5 {\vec{\tau}} \psi)^2\} 
\label{lag_njl}
\end{eqnarray}
where $m$ is the quark mass matrix, $m={\rm diag}(m_u,m_d)$, $g$ is
an effective coupling constant, and ${\vec{\tau}}$ represents the isospin 
Pauli matrices.

We investigate the pion and sigma meson masses at finite
temperature and chemical potential in the dimensional regularization.
To find the meson mass we calculate its propagator via
summation of the bubble type (fermion loop) diagrams.
The propagator for the scalar, $G_s$, and pseudo-scalar channel, $G_5$,
in the leading order of the $1/N_c$ expansion is given by
\begin{equation}
G_{s,5}(p^2,\langle \sigma \rangle) 
= \frac{4g^2}{2g-\Pi_{s,5}(p^2)} ,
\label{prop}
\end{equation}
where the self-energy $\Pi_{s,5}(p^2)$ is
\begin{equation}
\Pi_{s,5}(p^2) =  4 i g^2 \int \frac{d^D k}{(2\pi)^D} 
  \mbox{tr} \left[\Gamma_{s,5} \frac{1}{\not k - m 
  - \langle \sigma \rangle} \Gamma_{s,5} \frac{1}{(\not k - \not p)
  - m - \langle \sigma \rangle} \right] ,
\label{self}
\end{equation}
with $\Gamma_{s,5} = (\Gamma_s, \Gamma_5) = (1, i\gamma_5)$.
We regularize the self-energy (\ref{self}) by taking the space-time
dimensions as a parameter ($2<D<4$).
After having renormalized the coupling constant \cite{Inag:08}, we 
calculate the pion and sigma meson masses as the solutions of
the equations
\begin{eqnarray}
G_5^{-1}(p^2=m_\pi^2,\langle \sigma \rangle) = 0,
\quad {\rm Re} [G_s^{-1}(p^2=m_\sigma^2,\langle \sigma \rangle)] = 0 ,
\label{mass}
\end{eqnarray}
using the gap equation. These are not applicable in some cases,
for example, high temperature and/or large chemical
potential. In such cases we use
\begin{equation}
  \frac{\partial}
  {\partial p} |G_{s,5}(p^2=m_{\sigma,\pi}^2, \langle\sigma\rangle)| = 0 .
\label{mass2}
\end{equation}

\begin{figure}[t]
\begin{minipage}{69mm}
\begin{center}
\vglue 1mm
\includegraphics[width={!},height={40mm}]{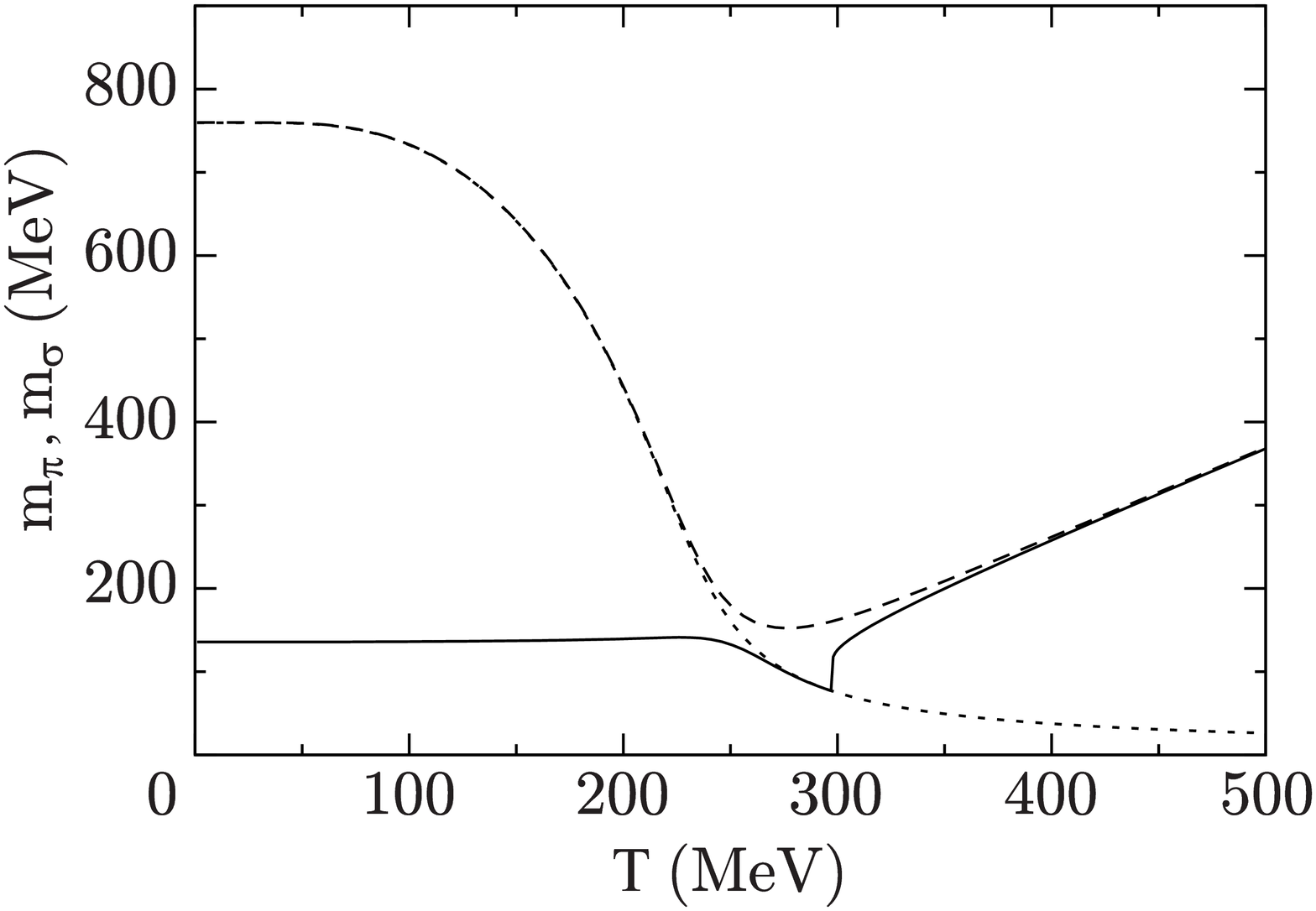}
\end{center}
\hspace{5mm}
{\footnotesize \hspace{5mm} (a) Pion and sigma meson masses.}
\vglue 5mm
\end{minipage}
\begin{minipage}{69mm}
\begin{center}
\vglue 1mm
\includegraphics[width={!},height={40mm}]{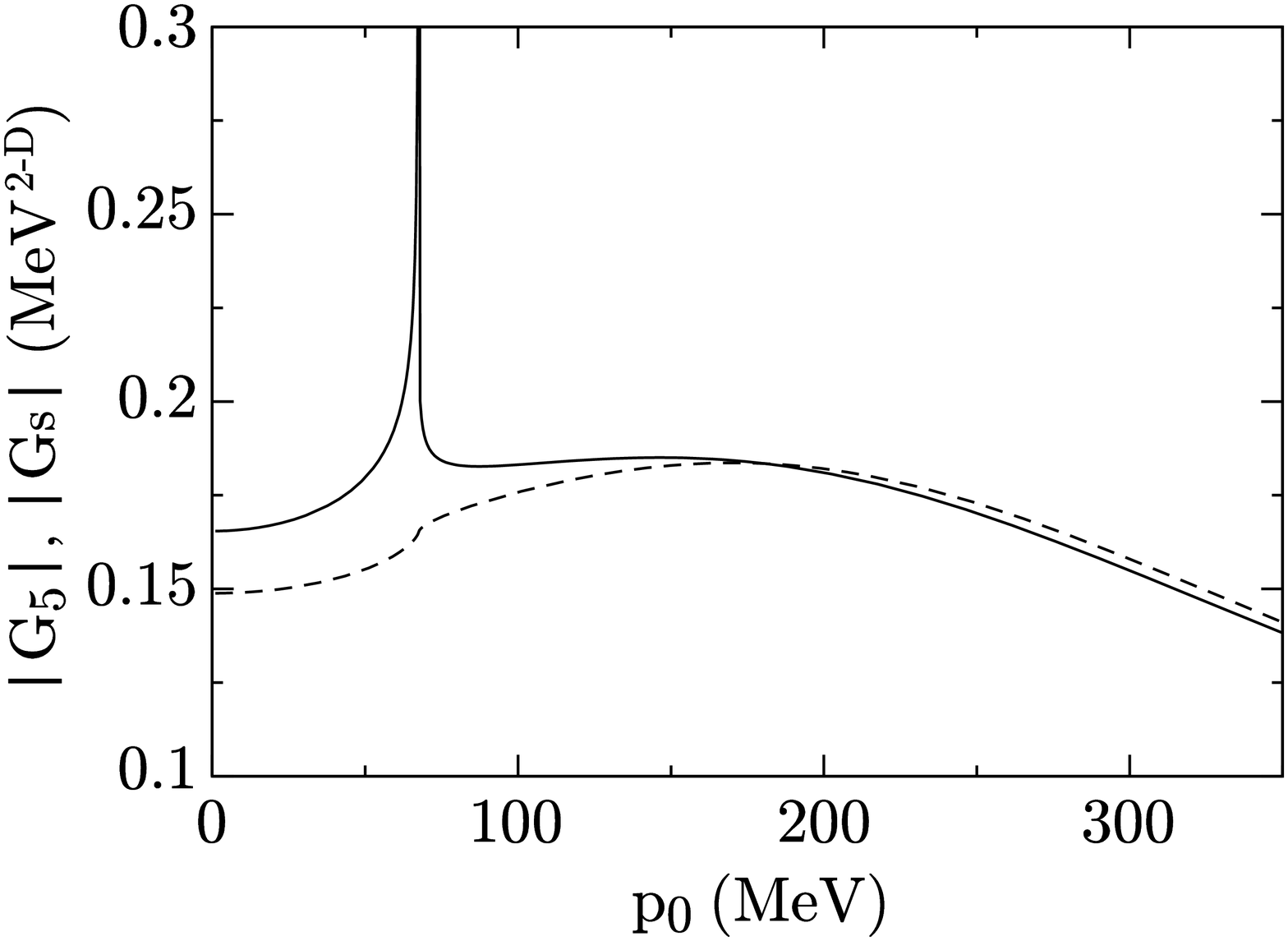}
\end{center}
\hspace{5mm}
{\footnotesize \hspace{5mm} (b) $|G_5|$ and $|G_s|$ at $T=310$ MeV.}
\vglue 5mm
\end{minipage}
\caption{(a) Pion and sigma meson masses as functions of the temperature
 at $\mu=100$ MeV.
 Full line represents the pion channel, dashed line the sigma
 channel, and dotted line $-2(\langle \sigma \rangle + m)$.
 (b) Meson propagator as functions of $p_0$ at $\mu=100$ MeV. 
 Full line represents 
 the pion propagator $|G_5|$, dashed line the sigma meson propagator $|G_s|$.}
\label{fig1}
\end{figure}

We use the imaginary time formalism, to introduce the temperature
and the chemical potential.
At the finite $T$ and $\mu$, the denominator of the fermion propagator 
in Eq. (\ref{self}) changes from $\not k - m - \langle \sigma \rangle$ to
$k_i \gamma_i - (\omega_n -i\mu)\gamma_4 + m + \langle \sigma \rangle$,
$\omega_n = (2n+1)\pi T$. The pion and sigma meson masses are obtained
by solving Eqs. (\ref{mass}) or (\ref{mass2}). To evaluate the meson
masses at finite $T$ and $\mu$, we fix the value of the model parameters 
$g,\ D$ \footnote{One degree of freedom still remains. We choose 
$D=2.4$ to evaluate the meson masses.} 
and renormalization scale by calculating
the pion mass and decay constant with $m_u = m_d = 5$ MeV  
in the dimensional regularization at $T=\mu=0$ \cite{Inag:08}.
In Fig. \ref{fig1} (a)
we plot the pion and sigma meson masses as functions of the temperature
at $\mu = 100$ MeV. The soft mode of the pion and sigma meson channels
almost degenerate above $T \simeq 300$ MeV. We also draw the behavior
of $|G_5|$ and $|G_s|$ at $\mu = 100$ MeV and $T = 310$ MeV in 
Fig. \ref{fig1} (b). The sharp peak structure of pion channel
corresponds to the term $-2(\langle \sigma \rangle + m)$ which is shown by 
the dotted line in Fig. \ref{fig1} (a).

\section{Phase structure}
At some large chemical potential a color Cooper pair of quarks
can be created in the $\bar3$ irreducible representation
of the $SU(3)$ color symmetry, 
i.e. color superconductivity phase transition takes place. 
To study the color superconducting phase the NJL model is extended to 
include the extra interaction term which is written in the 
diquark channel for the convenience. The model is given by
\begin{eqnarray}
 \mathcal{L} &=& \mathcal{L}_{\rm NJL}
    + g_D(\bar{\psi}^C_{ai}\varepsilon_{ij}
    \epsilon^b_{ac} i\gamma_5\psi_{c j})
   (\bar{\psi}_{d k}\varepsilon_{kl}\epsilon^b_{d e} i\gamma_5\psi^C_{e l}) ,
\label{lag}
\end{eqnarray}
where the indices $a,b,c,\cdots$ and $i,j,k,\cdots$ denote the 
color(1,2,3) and flavor($u,d$). 
$\psi^C$ represents the charge conjugate of $\psi$ and $g_D$ is the 
effective coupling constant for the diquark channel.

For practical calculations it is more convenient to introduce
the auxiliary fields, scalar $\sigma$, pseudo-scalar ${\vec{\pi}}$ 
and diquark $\Delta^b \sim -G_D \bar{\psi}^C\varepsilon\epsilon^b
i\gamma_5\psi$. For the simplicity, we take the massless quark limit 
and $\vec{\pi}=0$ owing to the chiral $SU(2)$ symmetry. 
The color $SU(3)$ symmetry allows us to set 
$\Delta^{1,2} = 0$ and $\Delta^3 = \Delta$.

To examine the phase structure of the extended NJL model
we evaluate the effective potential at finite $T$ and $\mu$.
\begin{eqnarray}
 V_{\rm eff}(\sigma,\Delta)&=&\frac{\sigma^2}{4g}+\frac{\Delta^2}{4g_D}
 - 
  \tilde{A} \int^\infty_0
  dp \ p^{D-2}
  \left[
   E_+ + E_- + E 
  +2T\ln(1+e^{-E_+/T}) \right. \nonumber \\
 && \left.
   +2T\ln(1+e^{- E_-/T}) 
   +T\ln(1+e^{- \xi_+/T})
   +T\ln(1+e^{- \xi_-/T})
  \right], 
\label{pot}
\end{eqnarray}
with
$E \equiv \sqrt{p^2 + \sigma^2}, 
 \xi_{\pm} \equiv (E \pm \mu), 
  E_{\pm} \equiv \sqrt{\Delta^2 + \xi_{\pm}^2}$ and
$\tilde{A} \equiv 4\sqrt{\pi} / [(2\pi)^{D/2} 
\Gamma\left(\frac{D-1}{2}\right)]$.
The difference of the effective potential (\ref{pot}) from
one derived in the cut-off regularization ($D=4$) 
is only in the third 
term of the right hand side
\begin{eqnarray}
\tilde{A}\int^\infty_0 dp \ p^{D-2}
\quad \to  \quad \frac{2}{\pi^2}\int^\Lambda_0 dp \ p^{2} ,
\label{dim-cut}
\end{eqnarray}
where $\Lambda$ is a cut-off scale.

\begin{figure}[t]
\begin{minipage}{69mm}
\begin{center}
\vglue 1mm
\includegraphics[width={!},height={40mm}]{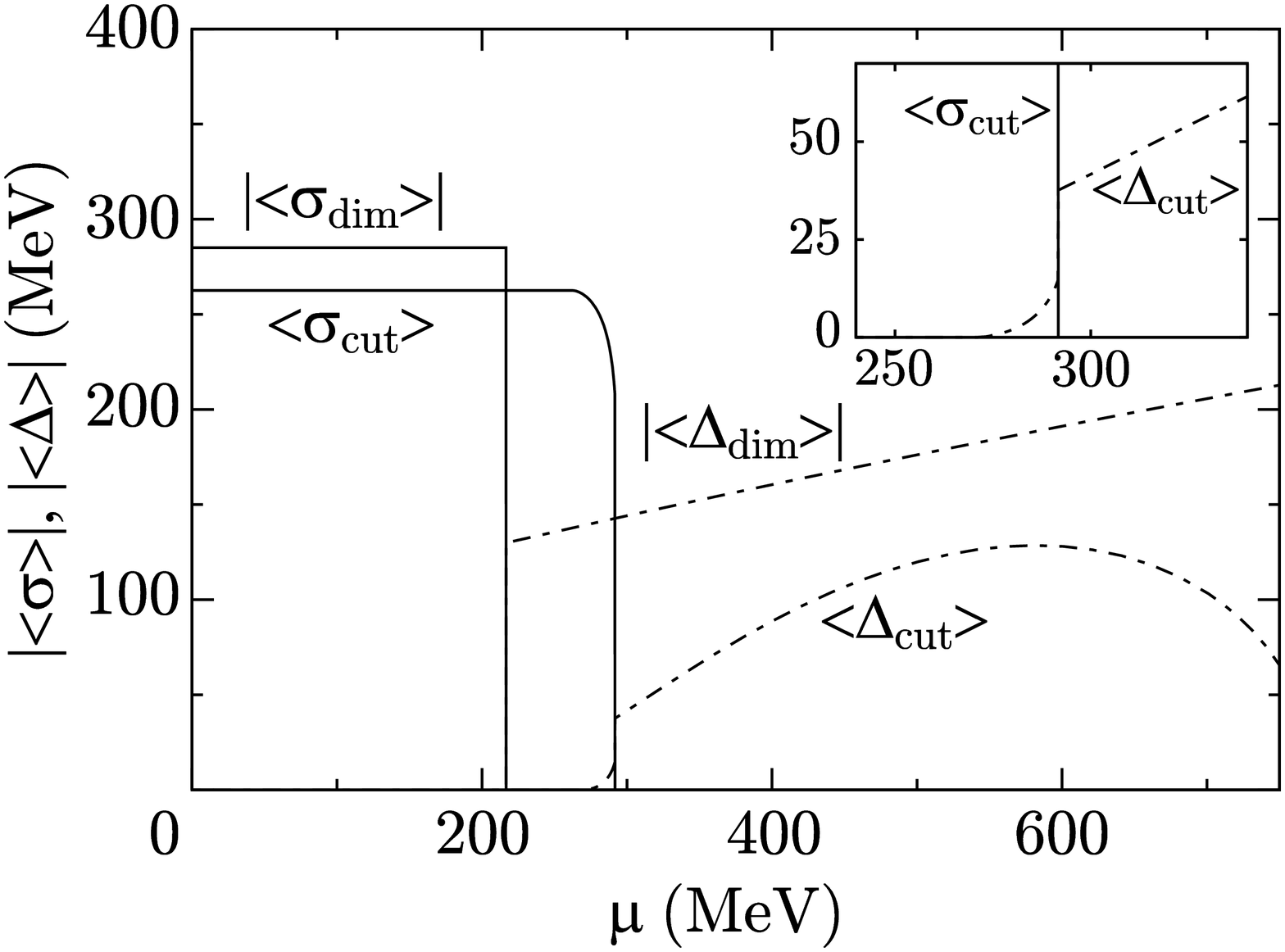}
\end{center}
\hspace{5mm}
{\footnotesize \hspace{5mm} (a) Solution of the gap equation. }
\vglue 5mm
\end{minipage}
\begin{minipage}{69mm}
\begin{center}
\vglue 1mm
\includegraphics[width={!},height={40mm}]{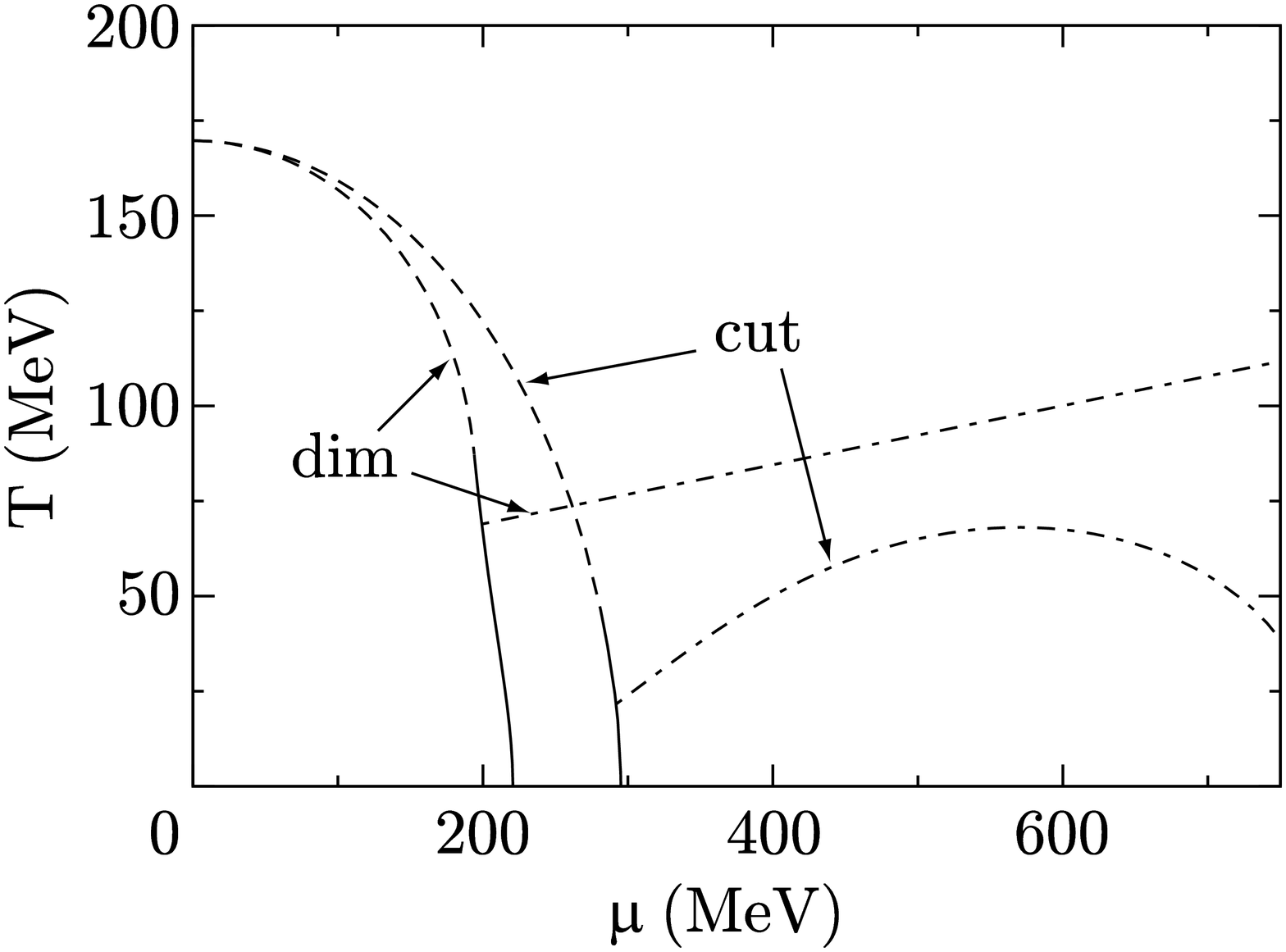}
\end{center}
\hspace{5mm}
{\footnotesize \hspace{5mm} (b) Phase diagram in $T$-$\mu$ plane.}
\vglue 5mm
\end{minipage}
 \caption{(a) Typical behavior of the gap equation's solution, 
$|\langle \sigma \rangle|$ and $|\langle \Delta \rangle|$  at $T\simeq 0$. 
Solid and dashed-dotted line represent 
$|\langle \sigma \rangle|$ and $|\langle \Delta \rangle|$ respectively.
(b) Dashed and solid lines indicate the 
second and first order phase transition of the chiral symmetry
breaking,
dashed-dotted line the second order phase transition of the 
color symmetry breaking.
}
\label{fig2}
\end{figure}

Using the effective potential (\ref{pot}) with Eq. (\ref{dim-cut}), 
we obtain the solution of the gap equation in Fig. \ref{fig2} (a)
and the phase structure of the chiral and color symmetry in 
Fig \ref{fig2} (b). \footnote{We fix the parameters in 
the cut-off($\Lambda=720$ MeV) and dimensional regularizations
to reproduce the pion mass and decay constant with 
$m_u = m_d = 4.5$ MeV. We set $D=2.28$ in order to get the 
same order critical temperature at $\mu=0$ as in the 
cut-off regularization.}
$\langle \Delta \rangle$ in the cut-off regularization decreases
near the cut-off scale in Fig. \ref{fig2} (a).
There is not coexistence phase of $\langle \sigma \rangle$ and 
$\langle \Delta \rangle$ in the dimensional regularization. 
Color superconducting phase in the cut-off regularization
also goes down near the cut-off scale in Fig. \ref{fig2} (b).
The tricritical point is located 
$(T,\ \mu)=(87.2,\ 194)$ and (46.7, 281) for the dimensional and cut-off
respectively.
  
\section{Summary}
We have discussed the meson masses and the phase structure of the 
two-flavor NJL model at finite $T$ and $\mu$ using the dimensional
regularization. The behavior of the meson masses at small $\mu$
is similar to that of the cut-off theory. It is found that the 
symmetry breaking of the chiral and color symmetry become 
different as $\mu$ is increased. In this region
one can see a new physical aspect in the NJL model
using the dimensional regularization. 
A paper about
the susceptibility and the high density stars is in preparation.

\end{document}